\documentclass[lettersize,journal]{IEEEtran}
\usepackage{amsmath,amsfonts}
\usepackage{algorithm}
\usepackage{algorithmicx} 
\usepackage{algpseudocode}
\usepackage{array}
\usepackage[caption=false,font=normalsize,labelfont=sf,textfont=sf]{subfig}
\usepackage{textcomp}
\usepackage{stfloats}
\usepackage{url}
\usepackage{verbatim}
\usepackage{graphicx}
\usepackage{cite}
\usepackage{tabularx}
\usepackage{booktabs}
\usepackage{multicol}
\usepackage[table]{xcolor}
\makeatletter
\renewcommand{\ALG@name}{Algorithm}

\renewcommand{\thealgorithm}{\Roman{algorithm}}
\newenvironment{breakablealgorithm}
{
		\begin{center}
			\refstepcounter{algorithm}
			\hrule height.8pt depth0pt \kern2pt
			\renewcommand{\caption}[2][\relax]{
				{\raggedright\textbf{\ALG@name~\thealgorithm} ##2\par}%
				\ifx\relax##1\relax 
				\addcontentsline{loa}{algorithm}{\protect\numberline{\thealgorithm}##2}%
				\else 
				\addcontentsline{loa}{algorithm}{\protect\numberline{\thealgorithm}##1}%
				\fi
				\kern2pt\hrule\kern2pt
			}
		}{
		\kern2pt\hrule\relax
	\end{center}
}

\makeatother

\begin{document}

\title{Stampede Alert Clustering Algorithmic System Based on Tiny-Scale Strengthened DETR}

\author{
   \IEEEauthorblockN{Mingze Sun\IEEEauthorrefmark{1}, Yiqing Wang\IEEEauthorrefmark{2}, Zhenyi Zhao\IEEEauthorrefmark{3}}
   \IEEEauthorblockA{\IEEEauthorrefmark{1}Beijing Jiaotong University, Beijing, China}

\thanks{This paper was produced by Intelligent Information Processing Laboratory, School of Electronics and Information Engineering, Beijing Jiaotong University, April 5, 2024}}


\markboth{Intelligent Information Processing Laboratory, School of EIE, Beijing Jiaotong University}%
{Shell \MakeLowercase{\textit{et al.}}: A Sample Article Using IEEEtran.cls for IEEE Journals}


\maketitle

\begin{abstract}
A novel crowd stampede detection and prediction algorithm based on Deformable DETR is proposed to address the challenges of detecting a large number of small targets and target occlusion in crowded airport and train station environments. In terms of model design, the algorithm incorporates a multi-scale feature fusion module to enlarge the receptive field and enhance the detection capability of small targets. Furthermore, the deformable attention mechanism is improved to reduce missed detections and false alarms for critical targets\cite{zhu2021deformable}. Additionally, a new algorithm is innovatively introduced for stampede event prediction and visualization. Experimental evaluations on the PKX-LHR dataset demonstrate that the enhanced algorithm achieves an 34\%\ performance in small target detection accuracy while maintaining the original detection speed. 
\end{abstract}


\section{Introduction}
\IEEEPARstart{I}{n} recent years, there has been a surge in safety incidents such as crowding and stampedes, underscoring the pivotal role of crowd detection in safeguarding pedestrian safety\cite{kefan2019analysis}. To mitigate the recurrence of similar tragedies, the anticipation and detection of stampede incidents in densely populated environments have become pressing concerns, particularly in locales such as airports and transportation hubs. However, the practical process of detecting small targets still confronts a multitude of challenges. Within dense crowds, targets may become occluded by one another or obstructed by physical barriers, resulting in indistinct boundaries and amplifying the complexity of detection. Furthermore, the dimensions and spatial distribution of targets in crowded settings often exhibit considerable variability, contributing to performance disparities among target detection algorithms across different regions or scales\cite{wang2024dense}. Furthermore, due to the unique hardware conditions at airports and train stations, excessive computational load can result in poor model performance and significantly reduced detection efficiency.

Compared to traditional methods that involve manual feature selection and handcrafted classifier design\cite{yin2016review}, object detection methods based on deep learning can automatically learn feature representations from raw data\cite{ahmed2021survey} and achieve accurate detection and localization of objects through end-to-end training.  Addressing the aforementioned scenario, Carion et al. developed the end-to-end object detection model DETR based on Transformers\cite{carion2020endtoend}. Subsequently, Deformable DETR integrated deformable attention mechanisms into the Transformer architecture, leading to innovations and improvements. This adaptation demonstrates advantages when handling complex environments such as dense crowd scenes.

This paper aims to improve the detection of small objects in dense crowds based on Deformable DETR and to develop a novel algorithm for predicting stampede incidents in crowded areas such as airports and train stations. The goal is to mitigate the occurrence of stampede incidents through these advancements.

\section{Related Works}
\subsection{Feature Fusion}
Feature fusion integrates feature information from different levels and sources, enabling a more comprehensive understanding and representation of the data. It improves model performance and generalization ability by integrating feature information from multiple perspectives, and helps enhance detection accuracy and robustness. Skillful design and application of feature fusion mechanisms can leverage valuable information in the data, thereby enhancing model performance, especially in object detection tasks. The dynamic allocation of weights to different features using attention mechanisms is a common approach in feature fusion, facilitating more effective focus on crucial features.
\subsection{Multi-scale Deformable Multi-head Attention Module}
In DeformAttention, the query does not calculate attention weights with every position in the global key space as in self-attention. Instead, for each query, a subset of positions in the global key space is sampled using a self-learned approach to compute local positional attention weights, which are then used to calculate attention with local values. By weighting different regions during image generation, deformable attention enables more flexible and fine-grained image synthesis.

The DeformAttention mechanism focuses attention only on specific pixels of the image rather than all pixels. In the deformable attention mechanism, the model learns attention weights for different regions of the image and adjusts the image generation process based on these weights. This approach effectively reduces computational overhead, enabling faster convergence. Furthermore, the reduced time complexity allows for easy utilization of multi-scale feature maps for prediction. Consequently, deformable attention enhances model performance in image generation tasks by producing images that more accurately reflect input conditions and contextual information, aligning better with input conditions and desired outputs.

The Multi-scale Deformable Multi-head Attention (MSDeformAttention) module is an effective attention mechanism for processing image feature maps. It successfully applies deformable attention to multi-scale features, providing more flexible and accurate attention information when processing feature maps.
\subsection{Transformer-based DETR}
DETR (DEtection TRansformer) serializes image features, then passes them through a Transformer Encoder to obtain correlated features of different parts on the image. It designs a specified number of object queries, which along with the output features from the Transformer Encoder, enter a Transformer Decoder.  Finally, after passing through a Feed Forward Network for dimension conversion, it obtains target classification and localization results for no more than the specified number of object queries. DETR serves as an end-to-end learning framework with strong modeling capabilities for global context, good scalability, and adaptability to targets of different sizes, achieving significant breakthroughs and accomplishments in the field of object detection.
\subsection{Deformable DETR Target Detection Model}
Deformable DETR is an improved version of the DETR object detection model, which introduces deformable attention mechanisms to enhance detection performance. It is an efficient and fast-converging end-to-end object detector that transforms the object detection task into a Transformer sequence modeling problem, simplifying the model training and optimization processes. Its core lies in the multiscale deformable multi-head attention module, enabling the model to better handle deformations and pose variations of objects in detection, thereby improving detection accuracy.

Compared to other object detection models, DETR incurs higher computational complexity and longer computation time due to continuously updating attention weights, requiring more epochs to converge. On the other hand, deformable DETR effectively learns sparse spatial information of images by leveraging the concept of deformable convolutions. However, deformable convolutions lack the ability to model relationships between objects in an image. Combining deformable convolutions with transformers provides a promising solution to address this limitation.

DETR is constrained by the substantial computational requirements of self-attention, thus it can only utilize the output of the final layer of the backbone feature extractor as the serialized input. It lacks a multi-scale operation strategy, which makes it difficult for DETR to detect small objects effectively. Deformable DETR addresses these issues by introducing deformable attention mechanisms, abandoning dense attention through point-wise alignment, and instead selectively calculating motion offsets based on the features of the current point across feature maps at multiple scales in a manner similar to deformable convolutions. This approach aims to find K relevant points. By using only these K points to compute the attention map, the scale of the attention map is significantly reduced, accelerating the model's convergence speed. As a result, Deformable DETR exhibits improved performance in detecting and localizing small-sized target objects with greater accuracy.

\section{The Design, Intent, and \\ Limitations of the Templates}
The algorithm employed in this study utilizes DETR as the baseline network and introduces multi-scale deformable attention mechanisms to achieve clearer object boundaries and more precise detection\cite{zhu2021deformable}. Additionally, the activation function Swish was substituted to reduce training costs and expedite convergence. Moreover, the integration of the tiny-scale feature fusion module enhances the representation capacity of small-scale features\cite{lin2017feature}.

This paper aims to improve the detection of small objects in dense crowds based on Deformable DETR and to develop a novel algorithm for predicting stampede incidents in crowded areas such as airports and train stations. The goal is to mitigate the occurrence of stampede incidents through these advancements.

\begin{figure}[htbp]  
  \centering
  \includegraphics[trim={20 17 6 12},scale=0.65,clip]{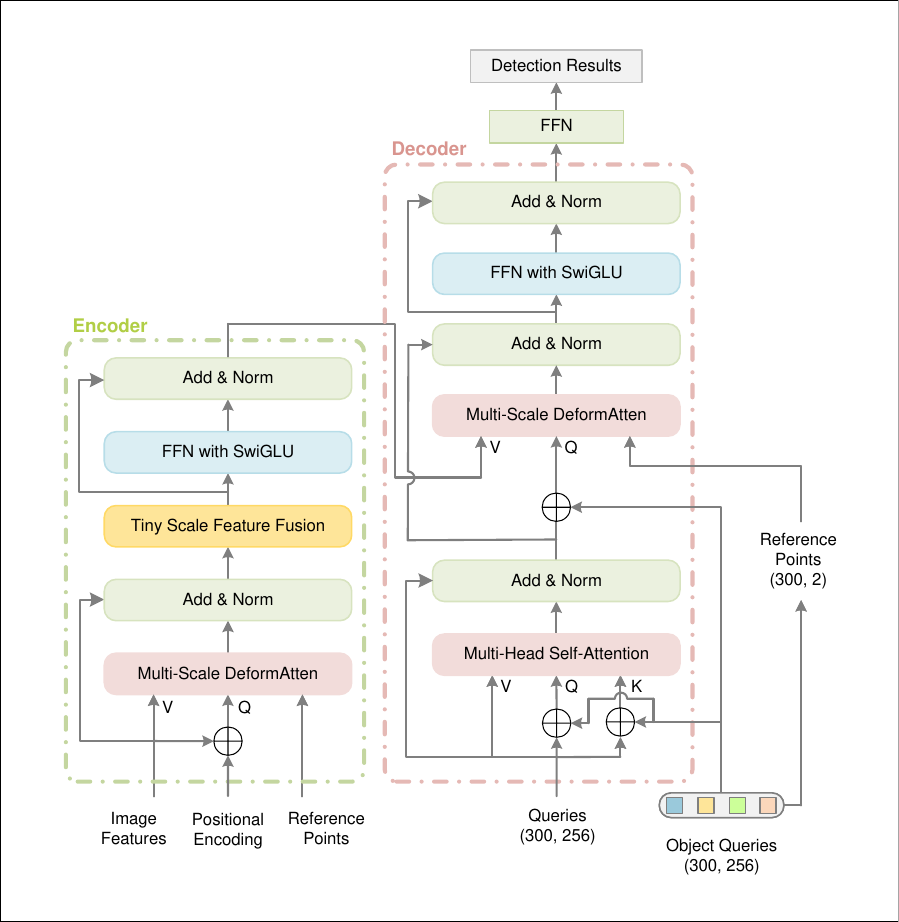}
  \caption{Architecture of Tiny-Scale Strengthened DETR}
  \label{fig:model}
\end{figure}
\subsection{The Core Architecture of the DETR Object Detection Model}
This article adopts the transformer as the underlying framework, replacing and adding new layers in the original model to achieve the task of small object detection.  It consists of an encoder and a decoder. Initially, the embedded representations of the source sequence and target sequence are combined with positional encodings as inputs to the encoder. Subsequently, these inputs are fed into both the encoder and decoder to generate bounding boxes and confidence scores. 
This article adopts the transformer as the underlying framework, replacing and adding new layers in the original model to achieve the task of small object detection.  It consists of an encoder and a decoder. Initially, the embedded representations of the source sequence and target sequence are combined with positional encodings as inputs to the encoder. Subsequently, these inputs are fed into both the encoder and decoder to generate bounding boxes and confidence scores.
The decoder part is also composed of multiple identical layers stacked together.  It consists of three sub-layers: Multi-Head Self-Attention, Multi-Scale DeformAtten, and FFN with SwiGlu.  Moreover, residual connections and layer normalization are employed within each layer.

\subsection{Introducing the SwiGLU Activation Function in FFN}

FFN (Feed-Forward Network) is a crucial element within the DETR model, employed to conduct further processing on features at each position within the Transformer encoder. Its mathematical expression is as follows:

\begin{equation}
\label{deqn_ex1a}
\text{FFN}(x) = \text{ReLU}(xW_1 + b_1)W_2 + b_2
\end{equation}

For the FFN layer with GELU activation, the mathematical representation is as follows:

\begin{equation}
\label{deqn_ex1a}
\text{GELU}(x) = x \cdot \Phi(x)
\end{equation}

Where \(Phi(x)\) is the Gaussian Error Linear Unit defined as:

\begin{equation}
\label{deqn_ex1a}
\Phi(x) = \frac{1}{2} \left( 1 + \text{erf} \left( \frac{x}{\sqrt{2}} \right) \right)
\end{equation}

In the FFN, the expression with GELU activation is:

\begin{equation}
\label{deqn_ex1a}
\text{FFN}_{\text{GELU}}(x) = \text{ReLU}(xW_1 + b_1)W_2 + b_2
\end{equation}

Here, x represents the input feature vector, \(W_1\) and \(b_1\) are the weight matrix and bias vector of the first fully connected layer, while \(W_2\) and \(b_2\) denote the weight matrix and bias vector of the second fully connected layer. In the FFN, the input undergoes a linear transformation (represented by \(xW_1 + b_1\)), followed by the application of the GELU activation function (denoted by \text{GELU}(x)), and finally, the output is obtained through the second linear transformation and bias term\cite{melaskyriazi2021need}.

The SwiGLU (Sigmoid-Weighted Linear Unit) is an activation function that offers several advantages over other common activation functions such as ReLU. Firstly, the SwiGLU function, possesses an unbounded upper limit, unlike ReLU, which lacks truncation. This characteristic enables SwiGLU to effectively handle larger input values without excessively compressing activation values. As a result, the SwiGLU function helps mitigate overfitting by maintaining a more suitable range for input values, thereby enhancing the model's generalization capability.

Secondly, the SwiGLU function exhibits enhanced regularization effects at the lower bound.  Its smooth profile and ubiquitous differentiability, in contrast to ReLU, contribute to sustained gradient continuity, facilitating smoother training dynamics and mitigating issues associated with gradient vanishing during backpropagation.  This inherent property endows the SwiGLU function with heightened resilience and steadfastness within intricate deep neural network architectures.

Additionally, the most significant difference between the SwiGLU and ReLU functions lies in their non-monotonic nature. For regions where the input values are less than zero, the SWiLU function gradually decreases, whereas for regions where the input values are greater than zero, the SWiGLU function gradually increases. This non-monotonicity can to some extent model distributions and better capture the nonlinear characteristics of input data, distinguishing it from both Swish and ReLU functions.

The expression for FFN with SwiGLU is shown below:

\begin{equation}
\label{deqn_ex1a}
Swish_{1} (xW_{1}+ b_{1})=(xW_{1}+ b_{1})\sigma (xW_{1}+ b_{1})
\end{equation}

\begin{equation}
\label{deqn_ex1a}
\small
SwiGLU(x,W_{1},V,b_{1},c )= Swish_{1} (xW_{1}+ b_{1})\otimes (xV+  b_{1})
\end{equation}

\begin{equation}
\label{deqn_ex1a}
\begin{split}
\tiny
FFN_{\text{SwiGLU}}(x,W_{1},V,b_{1},b_{2},c) =& \text{SwiGLU}(x,W_{1},V,b_{1},c)W_{2} \\
&+ b_{2}
\end{split}
\end{equation}

Equation (7) illustrates the computation process of the feedforward neural network based on SwiGLU.Here, the input is denoted as \(x\), and the parameters include weight matrices \(W_1\) and \(V\), bias vectors \(b_1\) and \(b_2\), and a parameter \(c\) controlling the shape of the Swish function. Firstly, the input \(x\) is processed by the SwiGLU activation function with parameters \(W_1\), \(V\), \(b_1\), and \(c\). Then, the result is multiplied by the weight matrix \(W_2\) and added with bias \(b_2\) to obtain the output of the network.

\begin{figure}[htbp]  
  \centering
  \includegraphics[trim={14 2 2 2},scale=0.21,clip]{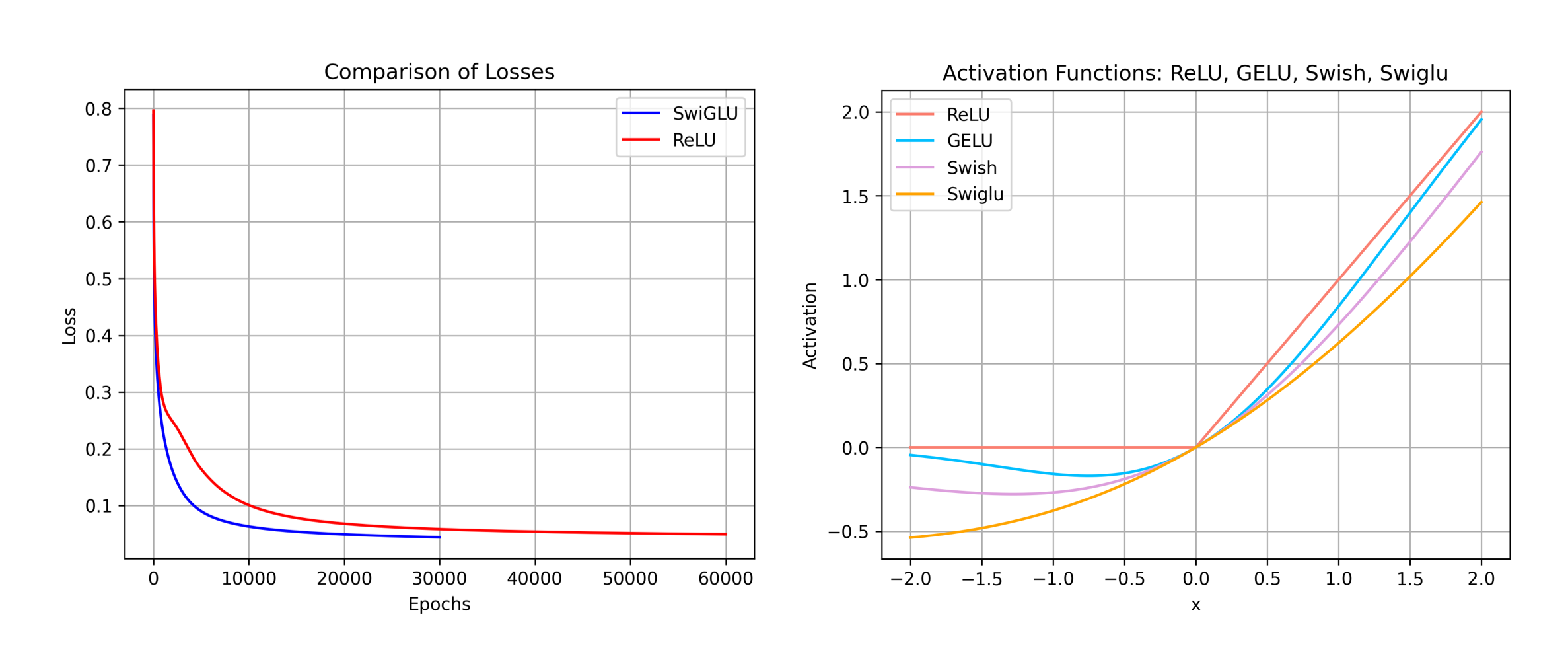}
  \caption{Comparison of several loss functions}
  \label{fig:model}
\end{figure}

To validate the superior performance of swiGLU over ReLU, we conducted experiments using feedforward neural networks for classification tasks, with a focus on visualizing decision boundaries (as depicted in Figure 2). Employing backpropagation for training, results show that using swiGLU leads to a 50 percent reduction in training time compared to ReLU, with lower final loss values upon convergence.

\begin{equation}
\label{eq:grad_a_L}
\frac{\partial \mathcal{L}}{\partial a^{(L)}}
\end{equation}

Next, we compute the gradients of the last layer and the biases:

\begin{equation}
\label{eq:grad_z_L}
\frac{\partial \mathcal{L}}{\partial z^{(L)}} = \frac{\partial \mathcal{L}}{\partial a^{(L)}} \odot g'(z^{(L)}) 
\end{equation}

\begin{equation}
\label{eq:grad_b_L}
\frac{\partial \mathcal{L}}{\partial b^{(L)}} = \frac{\partial \mathcal{L}}{\partial z^{(L)}} 
\end{equation}

Subsequently, we calculate the gradients of hidden layer $l$ and its parameters:

\begin{equation}
\label{eq:grad_a_l}
\frac{\partial \mathcal{L}}{\partial a^{(l)}} = \left( W^{(l+1)T} \frac{\partial \mathcal{L}}{\partial z^{(l+1)}} \right) \odot g'(z^{(l)}) 
\end{equation}

\begin{equation}
\label{eq:grad_W_l}
\frac{\partial \mathcal{L}}{\partial W^{(l)}} = \frac{\partial \mathcal{L}}{\partial z^{(l)}} (a^{(l-1)})^T 
\end{equation}

\begin{equation}
\label{eq:grad_b_l}
\frac{\partial \mathcal{L}}{\partial b^{(l)}} = \frac{\partial \mathcal{L}}{\partial z^{(l)}} 
\end{equation}

Finally, we update the parameters using the gradient descent optimization method or its variants:

\begin{equation}
\label{eq:update_W_l}
W^{(l)} := W^{(l)} - \alpha \frac{\partial \mathcal{L}}{\partial W^{(l)}} 
\end{equation}

\begin{equation}
\label{eq:update_b_l}
b^{(l)} := b^{(l)} - \alpha \frac{\partial \mathcal{L}}{\partial b^{(l)}} 
\end{equation}

It is worth noting that $W^{(l)T}$ represents the transpose of the weight matrix of layer $l$.

\begin{figure}[htbp]  
  \centering
  \includegraphics[trim={2 2 2 2},scale=0.26,clip]{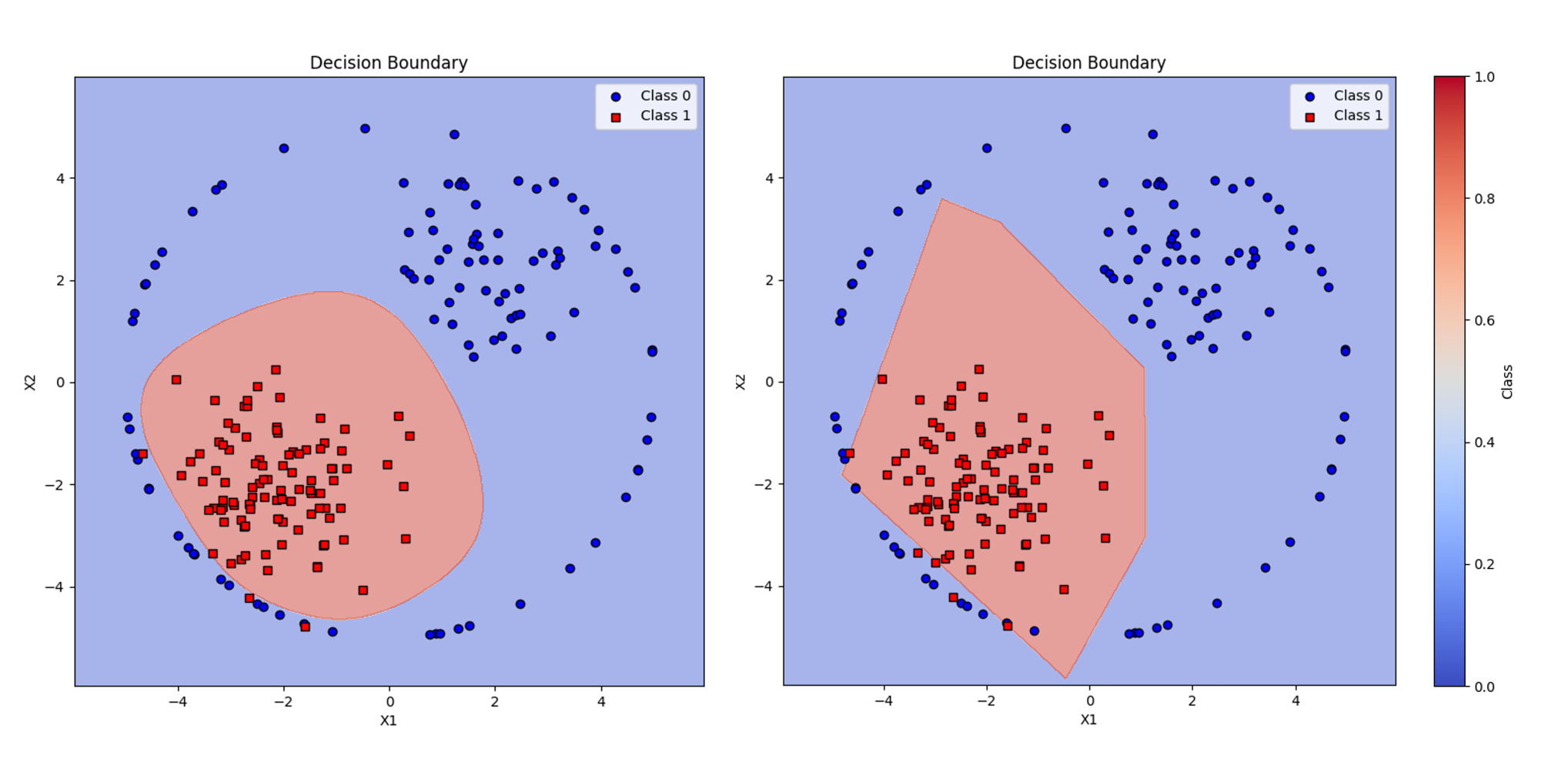}
  \caption{Decision boundary of the activation function}
  \label{fig:model}
\end{figure}


\subsection{Self-Attention}
In traditional attention mechanisms, only a fixed-size receptive field is typically considered for calculating attention weights. However, when dealing with images containing multi-scale information, a fixed-size receptive field may not effectively capture the relationships between details and global context. Therefore, the choice of Multi-Scale DeformAtten is made to address the processing of multi-scale information.

The self-attention mechanism is defined as follows:

\begin{equation}
\label{eq:attention}
\text{Attention}(Q,K,V) = \text{softmax}\left(\frac{Q \cdot K^T}{\sqrt{\text{dim}}}\right) \cdot V
\end{equation}

where the symbols denote: $Q$ - Query vector, $V$ - Value vector, $K$ - Key vector, $\text{dim}$ - input dimension.

 The attention mechanism calculates the compatibility between each query and key by taking the dot product of the query vector and the transpose of the key vector. Then, the dot product is divided by a scaling factor (the square root of the dimension) to keep the input to the softmax function within a reasonable range. The scaled dot product is passed through the softmax function to obtain attention weights relative to each query. Finally, the output values are weighted summed based on the attention weights to obtain the final attention output.

\subsection{Multi-Head Self-Attention}

Multi-head self-attention is computed as follows:

\begin{equation}
\label{eq:multi_head}
\text{MultiHead}(Q,K,V) = \text{Concat}(\text{head}_1, \dots, \text{head}_h) W^O
\end{equation}

where $\text{Concat}$ denotes the concatenation of outputs from multiple attention heads, $h$ is the number of heads, $\text{head}_i$ represents the output of the $i$-th head, and $W^O$ is the output weight matrix.

The output of each head $\text{head}_i$ is computed as:

\begin{equation}
\label{eq:head_i}
\text{head}_i = \text{Attention}(QW_i^Q, KW_i^K, VW_i^V)
\end{equation}

where $W_i^Q$, $W_i^K$, and $W_i^V$ are the query, key, and value transformation matrices for the $i$-th head, and $\text{Attention}$ is the self-attention calculation function.

\begin{figure}[htbp]  
  \centering
  \includegraphics[trim={4 4 4 8},scale=0.27,clip]{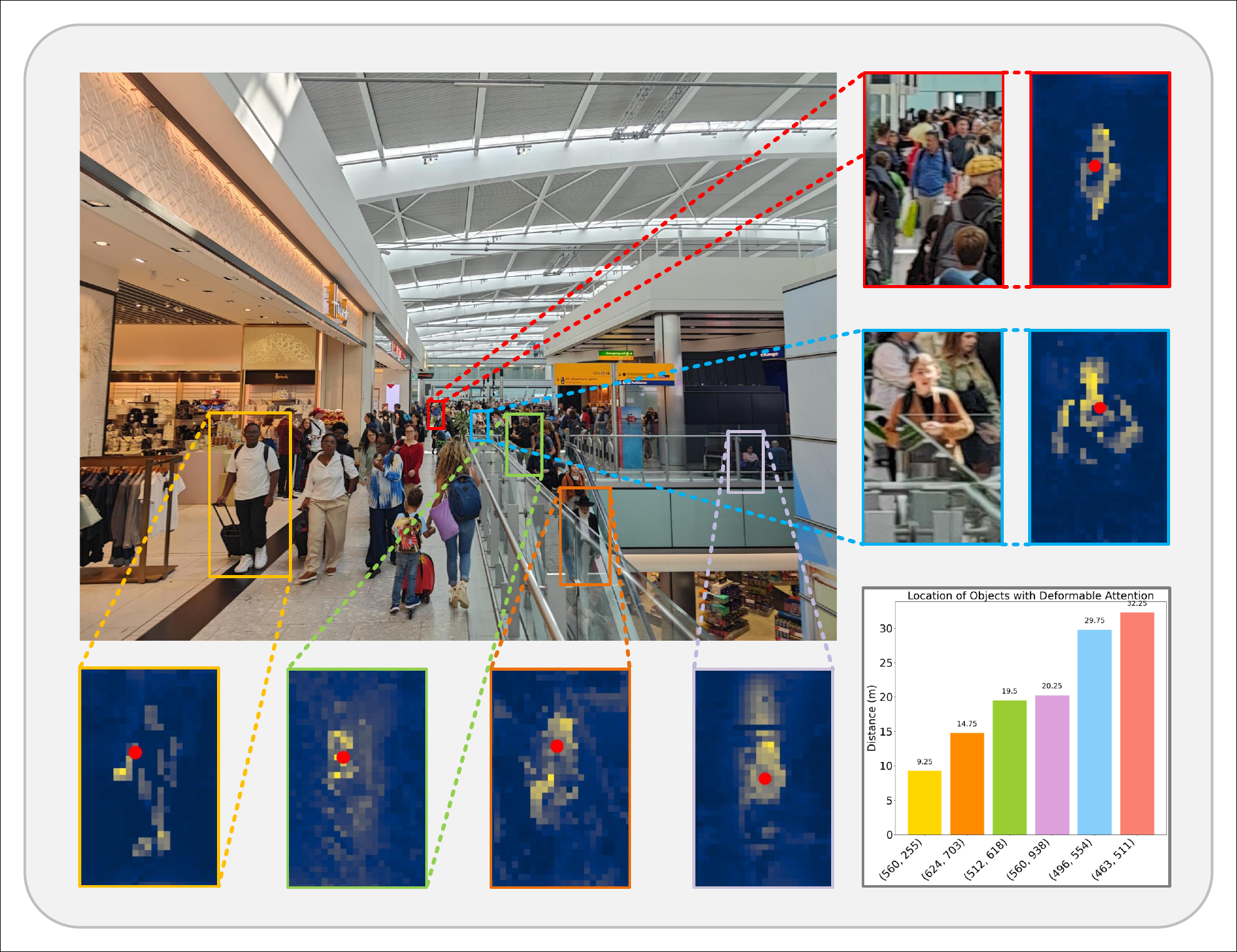}
  \caption{Multi-head attention mechanism based on PKSLHR data set}
  \label{fig:model}
\end{figure}

\section{The Training Process}
\subsection{Experimental environment and data set}
The main data sets are coco data set, tiny-person data set and self-built PKSLHR data set.The PKSLHR dataset comprises 0.52 gigabytes of images, which are partitioned into training and testing sets. This paper primarily utilizes the validation set from the dataset to describe the algorithm.

\subsection{Evaluation Criteria - IoU}
IoU (Intersection over Union) is employed to measure the degree of overlap between two bounding boxes or regions. It quantifies their similarity by computing the ratio of the intersection area to the union area of the two regions. The mathematical
representation is as follows:

\begin{equation}
\label{deqn_ex1a}
\text{IoU}=\frac{Area of Intersection of two boxes}{Area of Union of two boxes}
\end{equation}

Intersection area refers to the area of overlap between two regions, while "union area" refers to the total area when combining two regions. The value of IoU ranges from 0 to 1, where 0 indicates no overlap and 1 indicates complete overlap. Typically, in object detection tasks, when IoU is greater than a predefined threshold, it is considered a successful match between the predicted bounding box and the ground truth bounding box. By computing IOU between a set of predicted bounding boxes and ground truth bounding boxes, one can evaluate performance metrics such as accuracy and recall rate of object detection algorithms.

IoU Loss is the negative logarithmic form of Intersection over Union, the mathematical
representation is as follows:

\begin{equation}
\label{deqn_ex1a}
\text{IoU loss} = -\ln(IoU)
\end{equation}

It is commonly used in object detection tasks to measure the overlap between predicted bounding boxes and ground truth bounding boxes. Maximizing the IoU is employed to optimize the performance of object detection models.

\subsection{Experimental Results}

\begin{table*}[t]
  \centering
   \renewcommand{\arraystretch}{0.8}
  \caption{Object detection performance of different models}
 \begin{tabularx}{\textwidth}{l *{7}{>{\centering\arraybackslash}X}}
    \rowcolor{gray!30}
    \toprule
    \textbf{Model} & \textbf{Backbone} & \textbf{Epoch} & \textbf{AP} & \textbf{AP50} & \textbf{AP75} & \textbf{APs} & \textbf{APm} \\
    \midrule
    \textbf{DETR-DC5\cite{carion2020endtoend}} & Resnet-50 & 600 & 42.9 & 62.8 & 45.9 & 22.5 & 47.3 \\
                      & Resnet-101 & 675 & 44.2 & 63.7 & 46.7 & 23.8 & 48.1 \\
                      & Resnet-152 & 750 & 44.9 & 64.3 & 47.2 & 24.6 & 49.9 \\
                      & ResneXt-101 & 795 & 45.3 & 64.6 & 47.5 & 24.8 & 50.4 \\
    \midrule
    \textbf{Anchor-DETR-DC5\cite{liu2022dabdetr}} & Resnet-50 & 50 & 45.8 & 64.9 & 47.6 & 24.2 & 49 \\
                              & Resnet-101 & 65 & 47 & 65.7 & 48.5 & 25.5 & 49.9 \\
                              & Resnet-152 & 80 & 47.7 & 65.4 & 48.9 & 26.4 & 51.7 \\
                              & ResneXt-101 & 100 & 48.3 & 65.9 & 49.2 & 26.7 & 52.1 \\
    \midrule
    \textbf{SMCA-DETR\cite{9709993}} & Resnet-50 & 105 & 44.1 & 64 & 46.8 & 23.7 & 48.2 \\
                        & Resnet-101 & 120 & 45.2 & 64.8 & 47.7 & 24.8 & 49.1 \\
                        & Resnet-152 & 135 & 45.8 & 65.5 & 48.2 & 25.6 & 50.9 \\
                        & ResneXt-101 & 150 & 46.4 & 65.8 & 48.5 & 26.2 & 51.3 \\
    \midrule
    \textbf{Deformable-DETR\cite{zhu2021deformable}} & Resnet-50 & 55 & 46 & 64.9 & 49.8 & 28.7 & 49.4 \\
                              & Resnet-101 & 70 & 47.1 & 66.1 & 50.6 & 29.6 & 50.2 \\
                              & Resnet-152 & 85 & 47.7 & 66.6 & 51.1 & 30.1 & 50.7 \\
                              & ResneXt-101 & 100 & 48.1 & 66.9 & 51.5 & 30.4 & 51.3 \\
    \midrule
    \textbf{DN-Deformable-DETR\cite{10334480}} & Resnet-50 & 55 & 49.4 & 67.5 & 53.6 & 31.5 & 52.7 \\
                                 & Resnet-101 & 70 & 50.3 & 68.6 & 54.4 & 32.3 & 53.4 \\
                                 & Resnet-152 & 85 & 50.7 & 69.2 & 54.8 & 32.7 & 53.9 \\
                                 & ResneXt-101 & 100 & 51.4 & 69.5 & 55.2 & 33.1 & 54.2 \\
    \midrule
    \textbf{ST-Deformable-DETR (Ours)} & Resnet-50 & 45 & 50.7 & 67.9 & 54.1 & 32.2 & 52.6 \\
                                 & Resnet-101 & 55 & 51.5 & 68.8 & 55.2 & 33.1 & 53.5 \\
                                 & Resnet-152 & 65 & 51.8 & 69.4 & 55.7 & 33.6 & 55 \\
                                 & ResneXt-101 & 75 & 52.2 & 69.9 & 56.1 & 34 & 55.4 \\
    \bottomrule
 \end{tabularx}
  \label{tab:detection_performance}
\end{table*}

The training process involved evaluating different models with varying backbones and epochs to determine their performance based on Average Precision (AP) metrics at different Intersection over Union (IoU) thresholds (AP50, AP75) and for different object sizes (APs, APm, APl). Models including DETR-DC5, Anchor-DETR-DC5, SMCA-DETR, Deformable-DETR, DN-Deformable-DETR, and ST-Deformable-DETR were trained using backbone architectures such as Resnet-50, Resnet-101, Resnet-152, and ResneXt-101, for epochs ranging from 45 to 150. The goal was to identify the most effective configuration for end-to-end object detection tasks.The experimental results are shown in TABLE 1.

\subsection{Affine Transformation}

\begin{equation}
\label{eq:DFOV}
DFOV = 2 \times \arctan\left(\frac{\sqrt{w^2 + h^2}}{2f}\right)
\end{equation}

\begin{equation}
\label{eq:HFOV}
HFOV = 2 \times \arctan\left(\frac{w}{2f}\right)
\end{equation}

\begin{equation}
\label{eq:VFOV}
VFOV = 2 \times \arctan\left(\frac{h}{2f}\right)
\end{equation}

The general form of an affine transformation in 2D can be represented as:

\[
\begin{pmatrix}
x' \\
y' \\
1
\end{pmatrix}
=
\begin{pmatrix}
a & b & c \\
d & e & f \\
0 & 0 & 1
\end{pmatrix}
\begin{pmatrix}
x \\
y \\
1
\end{pmatrix}
\]

Where $(x, y)$ are the coordinates of a point in the original space, $(x', y')$ are the coordinates of the transformed point, and the transformation matrix is:

\[
\begin{pmatrix}
a & b & c \\
d & e & f \\
0 & 0 & 1
\end{pmatrix}
\]

$a$, $b$, $c$, $d$, $e$, and $f$ are the parameters of the transformation. $a$ and $e$ are responsible for scaling, $b$ and $d$ for shearing, and $c$ and $f$ for translation.

\begin{figure*}[htbp]
  \centering
   \includegraphics[trim={7 3 7 3}, scale=0.57, clip]{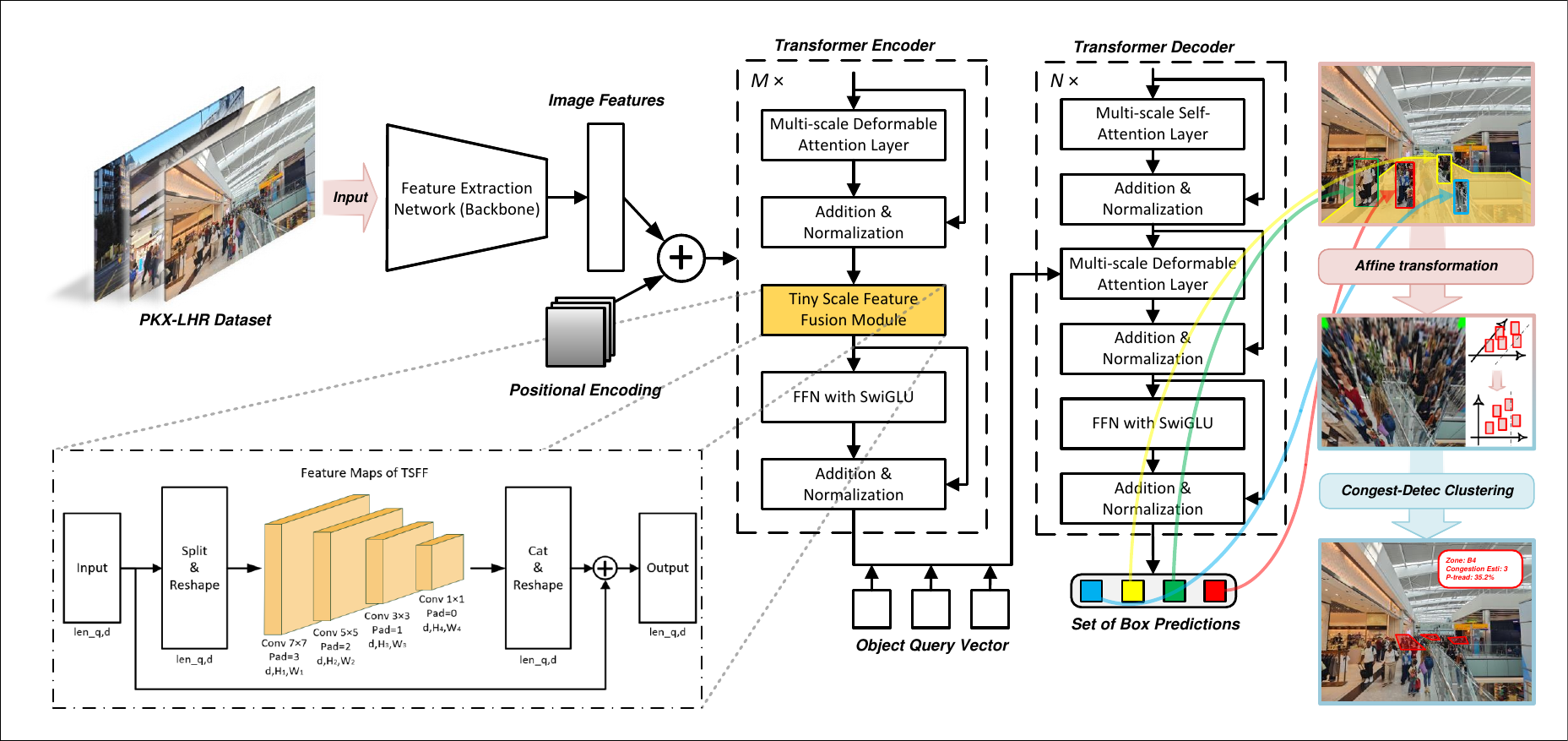}
  \caption{Algorithm overall frame diagram}
  \label{fig:your_label}
\end{figure*}  

\section{Crowded area detection clustering algorithm}
\subsection{Kmeans and KNN algorithms}
The K-means clustering algorithm is a widely used unsupervised learning method for partitioning a dataset into K distinct groups or clusters. Its objective is to minimize the total sum of distances between each sample and its assigned cluster centroid. Firstly, k samples are randomly initialized as the initial centers of mass of K clusters. Then all samples are assigned to the most similar cluster and the center of mass of each cluster is recalculated. Repeat the assignment process until the stop condition is met\cite{hua2016hybrid}. The K-means algorithm offers simplicity, computational efficiency, and scalability to large datasets. In the context of this study, the K-means clustering algorithm is proposed as a means to categorize the crowd, thereby laying the foundation for subsequent detection of crowd congestion.

K-nearest neighbors (KNN) clustering algorithm is a popular unsupervised learning technique utilized for data clustering and pattern recognition tasks. KNN assigns each data point to the cluster with the highest number of nearest neighbors, where the value of K represents the number of neighbors considered\cite{zhang2023fuzzy}. In this paper, the K-nearest neighbor (kNN) algorithm is modified to analyze crowd density by calculating the distance between k data points and setting a safety threshold to detect areas where stampedes may occur.
\subsection{Stampede prediction model based on kmeans and knn}
This section focuses on researching a semi-supervised congestion detection algorithm aimed at effectively preventing stampede incidents in densely populated areas such as airports and train stations using K-means and K-nearest neighbors algorithms\cite{hua2016fusion}.

Firstly, to obtain necessary data, the algorithm relies on security cameras, Wi-Fi signals, or other sensors to collect location information of individuals. This information will be used for subsequent analysis and processing. Crowd distribution is derived as the data input for the algorithm through small target detection. In the implementation of the algorithm, the K-means algorithm is used to cluster data points into a predetermined number of clusters. After initializing the cluster centers, each data point is assigned to the cluster whose center is closest to it based on the Euclidean distance\cite{Zahra2015}.Subsequently, the cluster centers are updated to the average of all data points within the cluster\cite{Abdulnassar2023}. This process continues until the cluster centers no longer change or reach a preset number of iterations.

The pseudo-code of this algorithm model is shown below.

\par

\begin{breakablealgorithm}
\caption{Crowd congestion detection algorithm}
\textbf{Input:} \\
\hspace*{1em} $DataPoint$: Population sample point size \\
\hspace*{1em} $k$: Number of center points in the crowd \\
\hspace*{1em} $K\_dist$: Distance calculation function (Euclidean Distance) \\
\hspace*{1em} $Point\_cent$: Population point distribution \\
\textbf{Output:} \\
\hspace*{1em} $Gather\_point$: Final crowd centers \\
\hspace*{1em} $Group\_assment$: Assignment of data points to crowd centers \\
\hspace*{1em} $Congestion\_degree$: Degree of crowd congestion
\begin{algorithmic}[1]
\State $m \gets \text{shape}(DataPoint)$
\State $Gather\_point \gets Point\_cent(DataPoint, k)$
\State $Group\_assment \gets \text{matrix of zeros with shape} \ (m, 2)$
\State $Group\_change \gets \text{True}$
\newpage
\While{$Group\_change$}
    \State $Group\_change \gets \text{False}$
    \For{$i \text{ in range}(m)$}
        \State $min\_K\_dist \gets \infty$
        \State $min\_class \gets -1$
        \For{$j \text{ in range}(k)$}
            \State $K\_distJI \gets K\_dist(Gather\_point[j, :], DataPoint[i, :])$
            \If{$min\_K\_dist > K\_distJI$}
                \State $min\_K\_dist \gets K\_distJI$
                \State $min\_class \gets j$
            \EndIf
        \EndFor
        \If{$Group\_assment[i, 0] \neq min\_class$}
            \State $Group\_change \gets \text{True}$
        \EndIf
        \State $Group\_assment[i, :] \gets min\_class, min\_K\_dist$
    \EndFor
    
    \For{$center\_i \text{ in range}(k)$}
        \State $Group \gets DataPoint[\text{nonzero}(Group\_assment[:, 0].A == center\_i)[0]]$
        \If{$Group.\text{size} > 0$}
            \State $Gather\_point[center\_i, :] \gets \text{mean}(Group, \text{axis}=0)$
        \EndIf
    \EndFor
\EndWhile

\State $Congestion\_degree \gets [ ]$

\For{$cent\_i \text{ in range}(k)$}
    \State $crowding\_distances \gets [ ]$
    \For{$i \text{ in range}(m)$}
        \If{$Group\_assment[i, 0] == cent\_i$}
            \State $distance\_iI \gets K\_dist(Gather\_point[cent\_i, :], DataPoint[i, :])$
            \State $crowding\_distances.\text{append}(distance\_iI)$
        \EndIf
    \EndFor
    \State $crowding\_distances.\text{sort}()$
    \State $c\_crowding\_distances \gets \text{crowding\_distances[:c]}$
    \State $count\_less\_than \gets \text{count}(dist \text{ for dist in } c\_crowding\_distances \text{ if } dist < Safe\_dist)$
    \If{$count\_less\_than > \frac{c}{Crowding\_factor}$}
        \State $Congestion\_degree.\text{append}(cent\_i)$
    \EndIf
\EndFor

\State $\textbf{return}$
\State $Gather\_point, \ Group\_assment, \ Congestion\_degree$
\end{algorithmic}
\end{breakablealgorithm}

After obtaining the distribution of crowd centroids and the cluster assignments of data points, the K-nearest neighbors algorithm is utilized to calculate the distances between data points and cluster centers. For each crowd centroid, the algorithm finds the c nearest data points to it. These distance calculations provide the basis for evaluating the congestion level. Evaluating congestion level is one of the key steps of this algorithm. In each iteration, by setting a safety distance and a congestion factor, the distances between each data point in a cluster and its cluster center are checked. Literature shows that five people per square meter are prone to or have been in stampedes. Therefore, 0.7m per capita distance is set as the safe distance in this paper\cite{huang2022intelligent}. If the distance between two points is less than this distance, it indicates that the two individuals are too close, possibly leading to a stampede. If the density of data points in a cluster exceeds the safety threshold, the cluster is considered to be in a congested state\cite{chen2023risk}.  This detection of congestion provides the basis for subsequent warnings and measures to be taken.

Finally, the algorithm labels the centroids of each group of people and the congestion situation in the image. Data points in congested states are outlined. Once a cluster is identified as congested, corresponding warning or dispatch measures will be taken to reduce congestion and prevent stampede incidents. This may include increasing security checkpoints, guiding pedestrian flow, reminding passengers to pay attention to safety, etc. The algorithm continuously iterates, updating cluster centers based on the latest data, and monitors congestion in real-time to ensure timely management of venue congestion and implementation of prevention measures. This research provides an effective method and tool for ensuring public safety and crowd management.

\begin{figure}[htbp]  
  \centering
  \includegraphics[scale=0.45,clip]{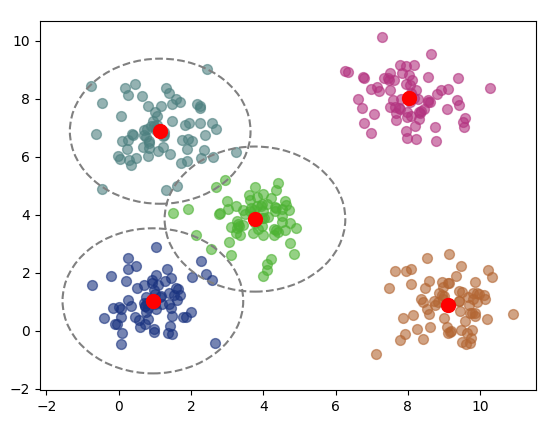}
  \caption{Detection result graph based on kmeans and knn clustering algorithm}
  \label{fig:model}
\end{figure}

The above images represent the visual output of the algorithm. Each point in the graph represents an individual, with different colors indicating the cluster to which they belong. The red circles denote the centroids calculated for each cluster. From the graph, it is evident that each clustered population is segmented into distinct regions, with their central positions marked. By computing the density of individuals within each cluster, areas prone to stampede incidents or where such incidents are occurring can be identified. The dashed boxes in the image highlight three densely populated areas, indicating the potential stampede risk associated with the corresponding population clusters. Thus, the algorithm appears to effectively address the detection needs for stampede incidents in crowded places, demonstrating promising feasibility.

\section{Conclusion}
The paper proposes a novel algorithmic model for detecting micro-scale stampede events based on an improved Deformable DETR framework combined with clustering algorithms. This algorithm exhibits superior overall detection accuracy as well as precision in detecting small targets. The contributions of this work primarily focus on three aspects:

(1) The algorithm utilized in this study is based on the DETR baseline network, with the incorporation of a multi-scale deformable attention mechanism aimed at achieving clearer object boundaries and more precise detection. Addressing the similarity in multi-scale perception capabilities of the original Deformable DETR, we propose a multi-scale deformable attention mechanism suitable for the Transformer-based object detection framework. This mechanism is applied to the Transformer encoder layers, thereby enhancing the efficiency and performance of feature fusion.

(2) In addressing the challenge of integrating small-scale features, we introduce a specially designed small-scale feature fusion module. This module is seamlessly integrated into our algorithm, aiming to further enhance the capability of capturing and representing micro features. This approach not only helps to increase the sensitivity of the target detection system to details but also enables more accurate capture of essential information such as object boundaries and textures, thereby effectively improving overall detection performance.

(3) In the Feed Forward Network (FFN), we have improved the activation function by adopting the SwiGLU activation function. This enhancement aims to reduce training costs, accelerate the convergence speed of the model, and further optimize the balance between cost and efficiency. By introducing the SwiGLU activation function, we have effectively enhanced the training efficiency of the neural network, enabling the model to better learn feature representations of the data within the same timeframe. Consequently, this leads to further improvements in the performance and practicality of the algorithm.

(4) The detection of congestion scenarios in this study employs a semi-supervised clustering algorithm combining k-means and k-nearest neighbors (knn). By computing the distances between individuals within crowds, it identifies clusters of people exceeding a predefined safety threshold and flags them accordingly. This approach provides a certain level of early warning for stampede incidents within crowds, offering a more reliable solution for enhancing safety measures in public spaces.

Future research will focus on the feature extraction capabilities of object detection models. We plan to conduct experiments on more datasets and evaluate their detection performance in practical scenarios.

\bibliographystyle{IEEEtran}
\bibliography{references}

\vfill

\end{document}